\documentclass[a4paper,fleqn]{cas-dc}

\usepackage[numbers,sort&compress]{natbib}
\usepackage{amsmath}
\usepackage{amssymb}
\usepackage{mathtools}
\usepackage{subfigure}
\usepackage{rotating}
\usepackage{amsmath}
\usepackage{amssymb}
\usepackage{mathtools}
\usepackage{multirow}
\usepackage{mwe}

\def\cuz{\mbox{CUPID-0}\xspace}

\def\mone{$\mathcal{M}_{1}$\xspace}

\def\radds{$^{226}$Ra\xspace}

\def\th{$^{232}$Th\xspace}
\def\raddq{$^{224}$Ra\xspace}
\def\rnddz{$^{220}$Rn\xspace}
\def\podus{$^{216}$Po\xspace}

\def\bbd{$2\nu\beta\beta$\xspace}
\def\onu{$0\nu\beta\beta$\xspace}

\def\Qbb{$Q_{\beta\beta}$}

\def\vita-dim{$T_{1/2}$\xspace}

\def\be{\begin{equation}}
\def\ee{\end{equation}}

\def\tpo{$T_{1/2}(^{216}\text{Po})$}

\begin{document}
\let\WriteBookmarks\relax
\def\floatpagepagefraction{1}
\def\textpagefraction{.001}
\shorttitle{Measurement of \podus half-life with the \cuz{} experiment}
\shortauthors{O.~Azzolini et~al.}

\title [mode = title]{Measurement of \podus half-life with the \cuz{} experiment}                      
\author[Legnaro]{O.~Azzolini}
\author[LBNL]{J.~W.~Beeman}
\author[Roma,INFNRoma]{F.~Bellini}
\author[MIB,INFNMiB,MattiaPA]{M.~Beretta}
\cormark[1]
\ead{mattia.beretta@berkeley.edu}
\author[INFNMiB]{M.~Biassoni}
\author[MIB,INFNMiB]{C.~Brofferio}
\author[LNGS]{C.~Bucci}
\author[MIB,INFNMiB]{S.~Capelli}
\author[INFNRoma]{L.~Cardani}
\author[MIB,INFNMiB]{P.~Carniti}
\author[INFNTorVergata,UNITorVergata,LNGS]{V.~Caracciolo}
\author[INFNRoma]{N.~Casali}
\author[MIB,INFNMiB]{D.~Chiesa}
\cormark[1]
\ead{davide.chiesa@mib.infn.it}
\author[MIB,INFNMiB]{M.~Clemenza}
\author[CNR,INFNRoma]{I.~Colantoni}
\author[INFNMiB]{O.~Cremonesi}
\author[INFNRoma]{A.~Cruciani}
\author[INFNRoma]{I.~Dafinei}
\author[GSSI,LNGS]{A.~D'Addabbo}
\author[Genova,INFNGenova]{S.~Di~Domizio}
\author[GSSI,INFNRoma]{F.~Ferroni}
\author[MIB,INFNMiB]{L.~Gironi}
\author[CNRS]{A.~Giuliani}
\author[LNGS]{P.~Gorla}
\author[INFNMiB]{C.~Gotti}
\author[Legnaro]{G.~Keppel}
\author[Roma,INFNRoma,MariaPA]{M.~Martinez}
\author[LNGS,GSSI,SergePA]{S.~Nagorny}
\author[MIB,INFNMiB]{M.~Nastasi}
\author[LNGS]{S.~Nisi}
\author[LNGS]{C.~Nones}
\author[LNGS]{D.~Orlandi}
\author[LNGS]{L.~Pagnanini}
\cormark[1]
\ead{lorenzo.pagnanini@lngs.infn.it}
\author[Genova,INFNGenova]{M.~Pallavicini}
\author[LNGS]{L.~Pattavina}
\author[MIB,INFNMiB]{M.~Pavan}
\author[INFNMiB]{G.~Pessina}
\author[INFNRoma]{V.~Pettinacci}
\author[LNGS]{S.~Pirro}
\author[MIB,INFNMiB]{S.~Pozzi}
\author[MIB,INFNMiB,LNGS]{E.~Previtali}
\author[GSSI,LNGS]{A.~Puiu}
\author[LNGS,USC]{C.~Rusconi}
\author[GSSI,LNGS]{K.~Sch\"affner}
\author[INFNRoma]{C.~Tomei}
\author[INFNRoma]{M.~Vignati}
\author[CNRS]{A.~Zolotarova}

\address[Legnaro]{INFN - Laboratori Nazionali di Legnaro, Legnaro (Padova) I-35020 - Italy}
\address[LBNL]{Materials Science Division, Lawrence Berkeley National Laboratory, Berkeley, CA 94720 - USA}
\address[Roma]{Dipartimento di Fisica, Sapienza Universit\`{a} di Roma, Roma I-00185 - Italy}
\address[INFNRoma]{INFN - Sezione di Roma, Roma I-00185 - Italy}
\address[MIB]{Dipartimento di Fisica, Universit\`{a} di Milano - Bicocca, Milano I-20126 - Italy}
\address[INFNMiB]{INFN - Sezione di Milano - Bicocca, Milano I-20126 - Italy}
\address[LNGS]{INFN - Laboratori Nazionali del Gran Sasso, Assergi (L'Aquila) I-67100 - Italy}
\address[CNR]{Consiglio Nazionale delle Ricerche - Istituto di Nanotecnologia, CNR -  Nanotec, Dip. Fisica, Sapienza Università di Roma, 00185 Roma, Italy}
\address[GSSI]{
Gran Sasso Science Institute, I-67100, L'Aquila - Italy}
\address[Genova]{Dipartimento di Fisica, Universit\`{a} di Genova, Genova I-16146 - Italy}
\address[INFNGenova]{INFN - Sezione di Genova, Genova I-16146 - Italy}
\address[CNRS]{
CSNSM, Universit\'e Paris-Saclay, CNRS/IN2P3, IJCLab, 91405 Orsay - France}
\address[CEA]{
IRFU, CEA, Universit\'e Paris-Saclay, F-91191 Gif-sur-Yvette, France}
\address[USC]{
Department of Physics and Astronomy, University of South Carolina, Columbia, SC 29208 - USA}
\address[INFNTorVergata]{INFN sezione di Roma “Tor Vergata”, I-00133 Rome, Italy}
\address[UNITorVergata]{Dipartimento di Fisica, Universit\`{a} di Roma “Tor Vergata”, I-00133, Rome, Italy}

\address[SergePA]{
Present Address: Queen's University, Physics Department, K7L 3N6, Kingston (ON), Canada}
\address[MariaPA]{
Present Address: Fundaci\'on ARAID and Laboratorio de F\'isica Nuclear y Astropart\'iculas, Universidad de Zaragoza, 50009 Zaragoza, Spain}
\address[MattiaPA]{
Present Address: Physics Department, University of California, Berkeley, CA 94720, USA}

\begin{abstract}
Rare event physics demands very detailed background control, high-performance detectors, and custom analysis strategies. 
Cryogenic calorimeters combine all these ingredients very effectively, representing a promising tool for next-generation experiments.  
\cuz{} is one of the most advanced examples of such a technique, having demonstrated its potential with several results obtained with limited exposure. 
In this paper, we present a further application. 
Exploiting the analysis of delayed coincidence, we can identify the signals caused by the \rnddz{}-\podus{} decay sequence on an event-by-event basis. 
The analysis of these events allows us to extract the time differences between the two decays, leading to a new evaluation of \podus{} half-life, estimated as $(143.3\pm2.8)$~ms.
\end{abstract}


\maketitle

\section{Introduction}
\label{Sec:Introduction}
Cryogenic calorimeters~\cite{Pirro:2017ecr} are a family of detectors increasingly adopted in experiments searching for rare events, such as dark matter interactions~\cite{Arnaud:2020svb,Abdelhameed:2019hmk,Angloher:2015ewa,Agnese:2017njq,Angloher:2017sft} or neutrinoless double beta decay~\cite{Alenkov:2019jis,Adams:2021rbc,Armengaud:2020luj,Azzolini:2019tta}.
These detectors, usually referred to as bolometers, are characterised by wide applicability, given the possibility to choose different materials as an active element.
In this context, scintillating calorimeters constitute a major improvement to this technique, providing particle identification through the analysis of phonon and light channels. 
The combination of different advantages makes this experimental technique a favourable option for future experiments \cite{app11041606}.
In recent years, cryogenic calorimeters have also proven their effectiveness performing precision measurement of two-neutrino double beta decay half-lives in the $10^{18}-10^{20}$~yr range~\cite{Azzolini:2019yib,Armengaud:2019rll,Adams:2021xiz}, relying on the global reconstruction of the measured energy spectrum. 
However, precision measurements of short (ms) nuclear lifetimes -- compared to the $>10$~ms response times of bolometric detectors -- based on the reconstruction of the time intervals between events are not a standard application for this technique.

In \cuz{}, a \onu{} experiment based on the scintillating bolometers technique~\cite{Azzolini:2019tta}, we combine $\alpha$-particle tagging and identification of delayed coincidences between subsequent decays to identify the different background sources contributing to the measured spectrum \cite{Azzolini:2019nmi,Azzolini:2021yft}.
This identification provided us with a high-purity sample of \raddq{}-\rnddz{}-\podus{} cascade, belonging to the \th{} chain. By developing dedicated processing for the \rnddz{}-\podus{} events, we were able to measure the \podus{} half-life (\vita-dim) with high precision, providing an additional reference value for this physical constant.

Since \podus{} half-life is still a debated subject \cite{216Po_Discussion_LNHB_CEA}, additional measurements can help clarifying the current scenario and obtaining a more accurate evaluation. Moreover, this work can pave the way to further studies aimed to investigate the claim that $\alpha$-decay half-live of nuclides embedded in a metal changes at cryogenic temperature~\cite{raiola_first_2007}.

\noindent
In section \ref{Sec:SetupAndData}, we briefly outline the \cuz{} experiment and the data production sequence. In section \ref{Sec:SearchDelayed}, we present the selection criteria used to identify delayed coincidences in the data. In section \ref{Sec:Results}, we detail the \vita-dim{} measurement and its uncertainty assessment. Lastly, in section \ref{Sec:Conclusions}, we combine the value obtained for \vita-dim{} with its previous measurements, quoting a new best evaluation.

\section{Experimental setup and data production}
\label{Sec:SetupAndData}
\cuz{} is the first \onu{} experiment implementing the simultaneous heat-light readout
with enriched scintillating bolometers~\cite{Azzolini:2018tum}.
Its main goal is the search for the \onu{} of $^{82}$Se (\Qbb = 2997.9 $\pm$ 0.3~keV~\cite{Lincoln:2012fq}). 

The detector is composed of 24 Zn$^{82}$Se crystals 95\% enriched in $^{82}$Se and two non-enriched ZnSe crystals, building up a total mass of 10.5~kg.
When a particle releases its energy in a ZnSe crystal, kept at a temperature of $\sim$10~mK, both a sizeable increase in temperature and scintillation light are produced.
To collect the latter signal, each crystal faces two germanium wafers (one on the top and one at the bottom) operated as calorimetric light detectors (LDs)~\cite{Beeman:2013zva}. 
Both ZnSe crystals and LDs are equipped with a germanium Neutron Transition Doped (NTD) thermistor, acting as temperature-voltage transducer~\cite{Haller}, to measure the energy deposit in the ZnSe crystal and the light absorbed in the LD. 
More details on the experimental setup can be found in Ref.~\cite{Azzolini:2018tum}.

The data used for this work have been collected in Phase I (June 2017 to December 2018) and Phase II (May 2019 to February 2020) of \cuz{}, and correspond to a total exposure of 15.74~kg~yr ($\sim$15776 h).

Both ZnSe crystals and LDs signals are filtered~\cite{Arnaboldi:2017aek} and then continuously digitized~\cite{DiDomizio:2018ldc} to be subsequently processed with an offline analysis chain.
An event is defined when a derivative trigger fires on the ZnSe crystals, identifying a 5~s waveform (1~s and 4~s of pre-, pos-trigger respectively).
The sampling frequency is 1000~Hz and 2000~Hz for the ZnSe and LDs, respectively.
Each event is processed with a matched-filter~\cite{Gatti:1986cw} to improve the signal-to-noise ratio. A full description of the analysis procedure is reported in~\cite{Azzolini:2018yye}. For the current analysis purpose, we recall that the average rise-time of the ZnSe signal, i.e the time interval between the 10 and 90\% of the leading edge of the pulse amplitude, is about 13.5~ms~\cite{Azzolini:2018tum}.

To refine the data quality, a series of selections is performed. 
In particular, events with a specific pulse shape are labelled as $\alpha$-particle-induced. The chosen criteria, defined in Ref.~\cite{Azzolini:2018yye}, allow us to tag $>$99.9\% of $\alpha$ events with energies above 2~MeV.
To further refine the selection, events occurring in different crystals with a time distance less than 40~ms are tagged with a multiplicity label ($\mathcal{M}_\#$), equal to the number of involved crystals (\#). 
Lastly, an event is labelled as pile-up if the derivative trigger fires more than once in the first 5~s window.

For the presented analysis only $\mathcal{M}_1$ data have been chosen.
Finally, events from bad behaving crystals (two enriched crystals and two natural crystals) are not considered.
As a consequence, the data we are presenting correspond to an active mass of 8.74~kg.

\section{Search for delayed coincidences}
\label{Sec:SearchDelayed}
The sequence of $\alpha$ decays of interest for this work is the following:
\begin{align*} 
^{224}\text{Ra}&\xRightarrow[3.66~\text{d}]{5.79~\text{MeV}}~^{220}\text{Rn}\xRightarrow[55.6~\text{s}]{6.4~\text{MeV}}~^{216}\text{Po}\xRightarrow[145~\text{ms}]{6.9~\text{MeV}}~^{212}\text{Pb}
\end{align*}
belonging to the lower part of \th{} chain. Given the time scale of \cuz{} pulses, the \podus decay (\vita-dim{} = $(145\pm2)$~ms \cite{Wu:2007cdl}) is likely to produce a pile-up event with the previous $^{220}$Rn decay.

We identify delayed $\alpha$-$\alpha$ coincidences combining $\alpha$-particle identification and time correlation between subsequent decays in a single crystal (\mone events).

When an $\alpha$ event with energy around the \raddq{} Q-value is detected (i.e. an event fully within the crystal volume), we open a coincidence window $\Delta t_w = 5 \cdot T_{1/2}$ of the \rnddz{} decay, equal to 278~s. In total we identified 4922 events belonging to this category, opening an equal number of coincidence windows. If an $\alpha$-event is detected in this time period, we label this event as a \rnddz{}-\podus{} pileup. 
From previous analyses on the \cuz{} background \cite{Azzolini:2019nmi}, we know that contamination are mainly localised in the bulk of the crystals and not on their surface. As a consequence, the whole decay sequence is most likely to be contained in a single crystal, with a Monte Carlo evaluated efficiency of 65.7\%~\cite{Azzolini:2021yft}.
The expected number of random coincidences polluting the selected events can be calculated and it is expected to be small. The probability of random $\alpha$-$\alpha$ coincidences in a 278~s window is $<4\cdot10^{-2}$, given the low $\alpha$ count rate of \cuz{} ($<1.7 \times 10^{-4}$ Hz/crystal)~\cite{Azzolini:2021yft}.
The probability for a random pile-up in the same event (4~s window) simulating a \rnddz{}-\podus{} sequence is $<8\cdot10^{-3}$ given the $\sim2$~mHz counting rate for each crystal, considering bot $\alpha$ and $\beta/\gamma$ events. The combined probability of a random triple coincidence is, assuming these values, $<3\cdot10^{4}$. Combining this information with the total number of initial coincidence windows, we expect $<1$ random coincidences.

\begin{figure*}
    \centering
    \begin{minipage}[c]{.4\textwidth}
    \includegraphics[width=\textwidth]{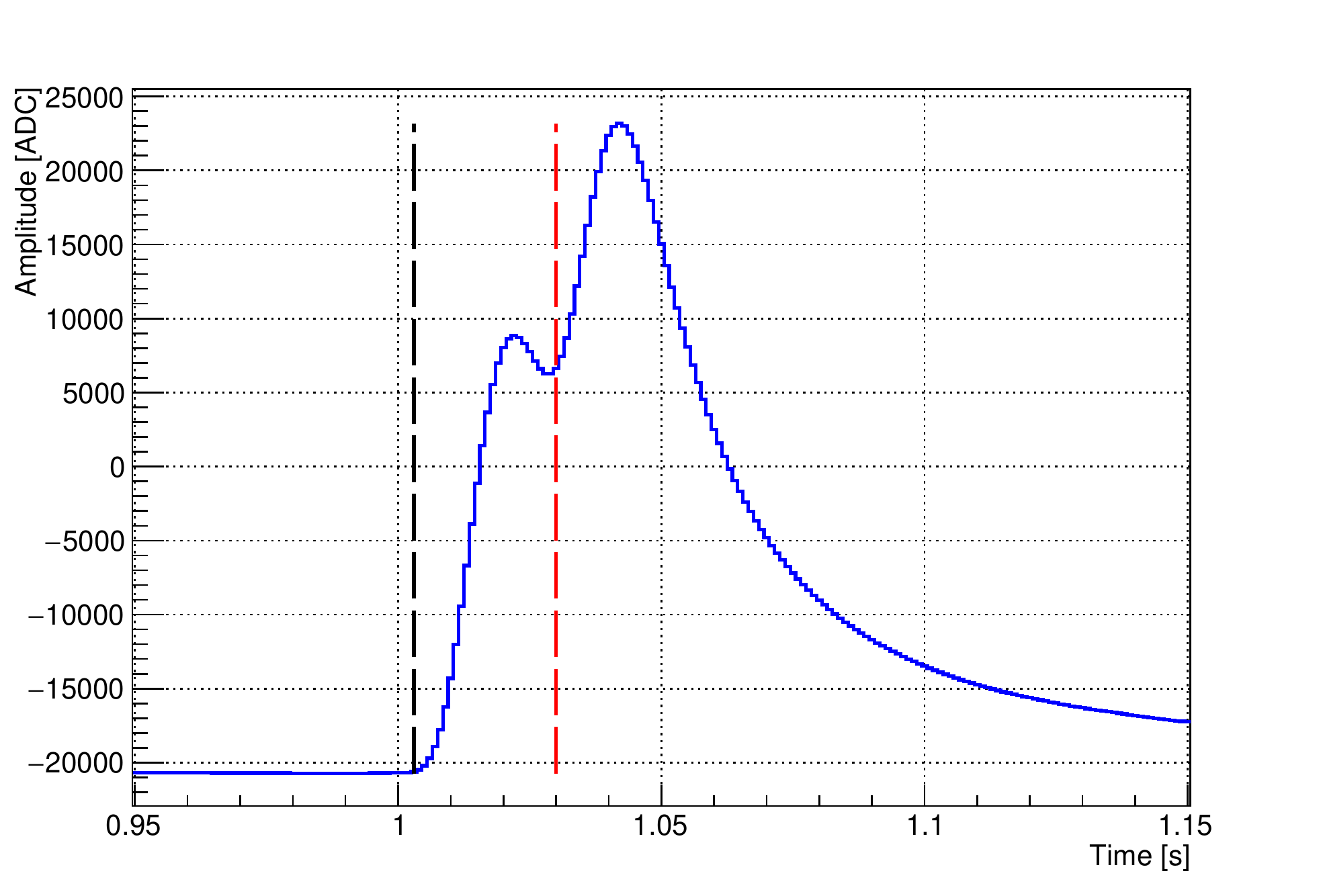}
    \end{minipage}
    \begin{minipage}[c]{.4\textwidth}
    \includegraphics[width=\textwidth]{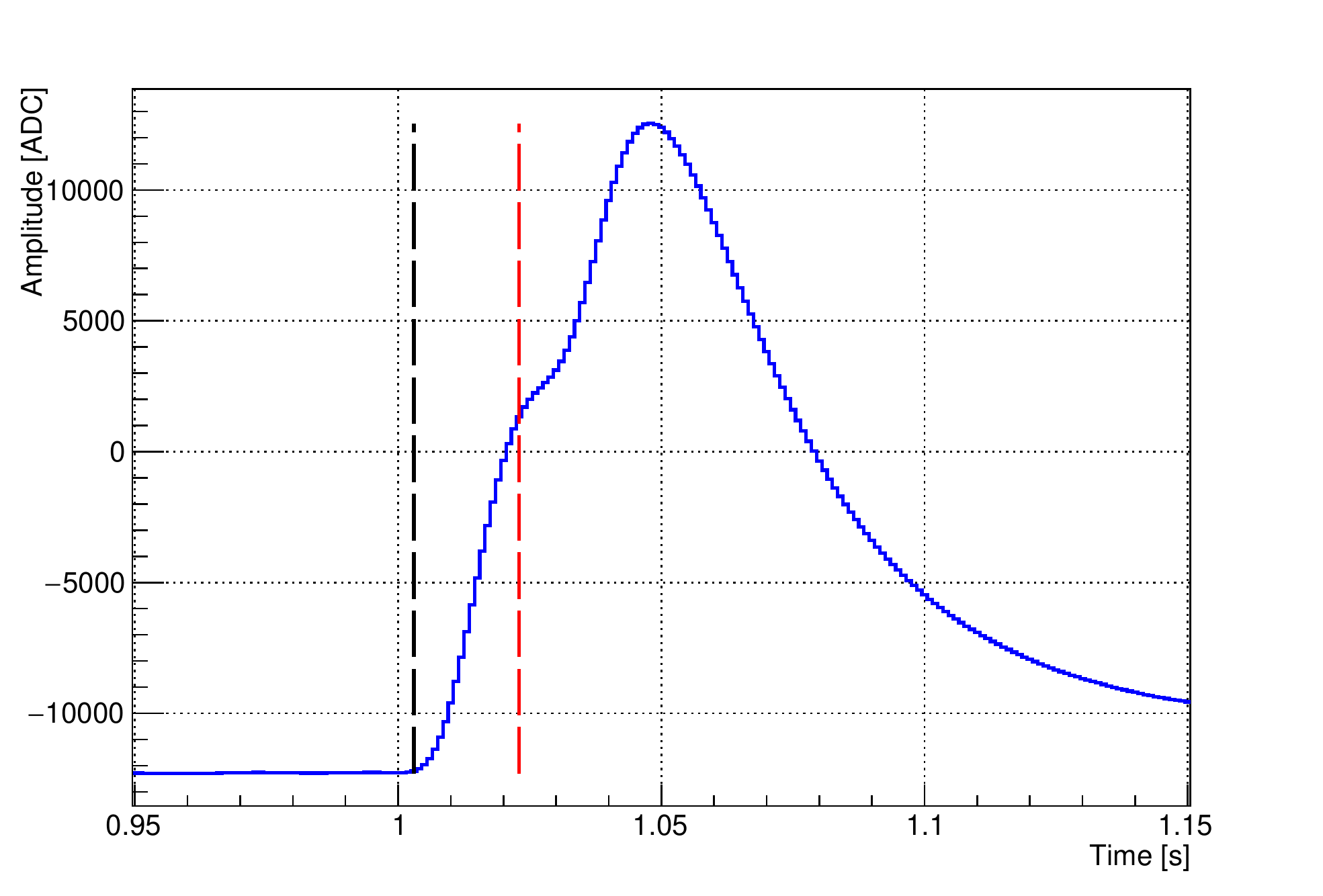}
     \end{minipage}
    \caption{Example of \rnddz{}-\podus{} pile-up events. The time position of the two pulses is identified with a derivative trigger tuned to maximise the time sensitivity at the signal-to-noise ratio of $\alpha$-events. If the two signals are spaced by more than 25~ms (left), the trigger positions are well identified (vertical dashed lines), while the uncertainty becomes large for signals with a time difference comparable with the pulse rise-time (right).}
    \label{fig:PileupExample}
\end{figure*}

We can therefore obtain a pure subset of \rnddz{}-\podus{} events with good efficiency, focusing our decay time analysis only on the interesting events. The selected events resemble the one reported in the left panel of Figure~\ref{fig:PileupExample}. 

To reconstruct the time difference in the pile-up doublet, we use a derivative trigger optimised to select $\alpha$ events. As can be guessed from the right panel of Figure~\ref{fig:PileupExample}, when the time difference between the two events approaches the rise-time of the pulse, the derivative algorithm may run into a higher error in the time reconstruction. To exclude these events from the analysis, we set a common minimum acceptable time difference of 25~ms, higher than all the different rise-times of \cuz{} crystals (cfr. \cite{Azzolini:2018tum}, table 4). This lower limit is a cautionary choice, so as to keep in the analysis only events where the reconstruction of the two pulses is reliable without the need for a channel-by-channel optimisation of the algorithm. Assuming \vita-dim{} = $(145\pm2)$~ms, an efficiency of 84.4\% can be calculated for this selection.

With this additional processing step, we can select among the tagged \rnddz{}-\podus{} pile-up events only those showing two distinct pulses. 
The total number of selected events is 2702. This value is compatible with the expected number of events, 2729$\pm$52, obtained combining the number of initial coincidence windows opened (4922) with the containment efficiency (65.7\%) and the time selection efficiency (84.4\%). 
We use the time difference calculated for these events to evaluate the half-life of the \podus{} $\alpha$-decay.

\section{Results}
\label{Sec:Results}
\begin{figure}[t!]
    \centering
    \includegraphics[width = 0.49\textwidth]{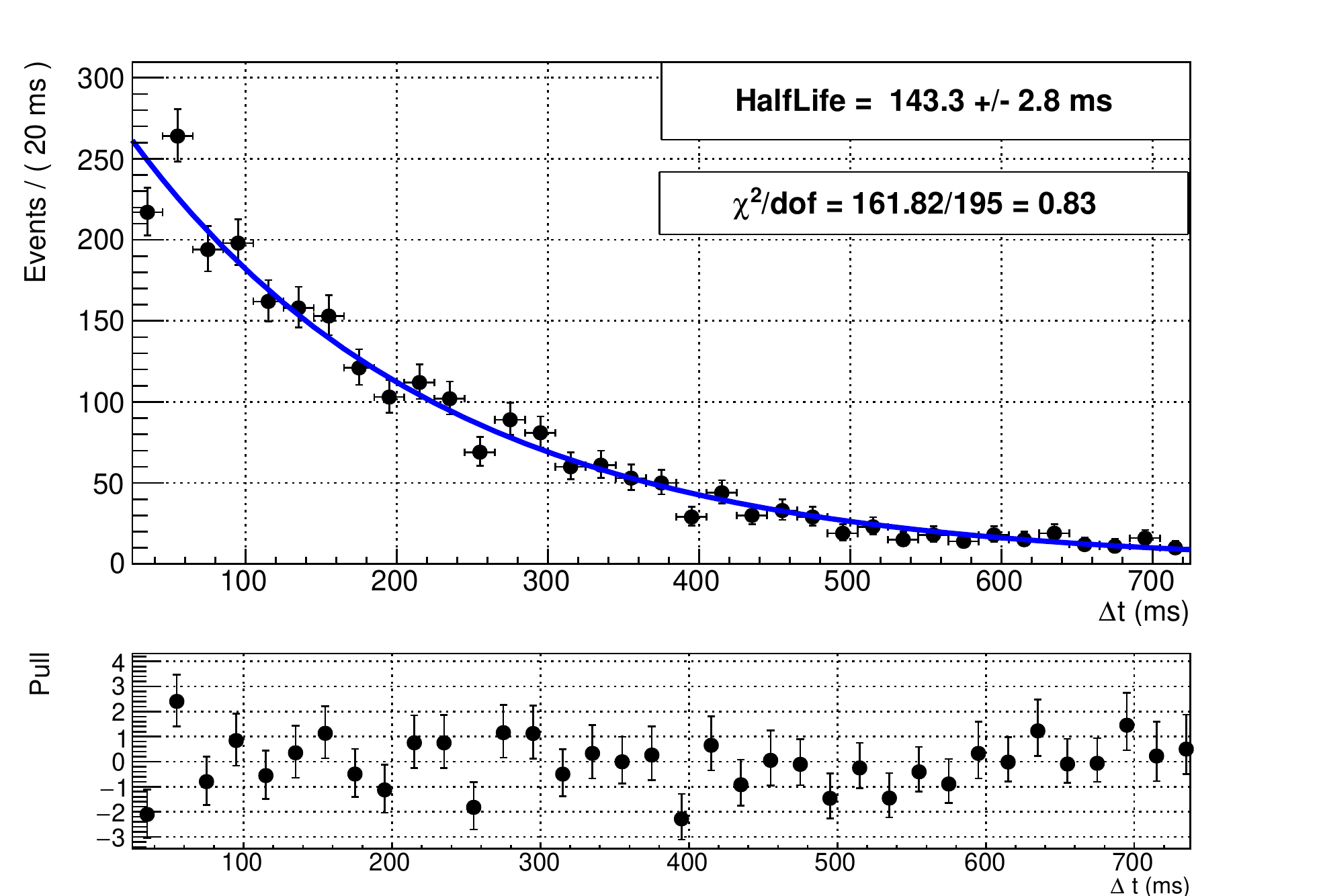}
    \caption{Time difference distribution of \rnddz{}-\podus{} events (top). The distribution is described with an exponential decay and a flat background to account for eventual random coincidences. We perform the fit with an unbinned extended maximum likelihood procedure. The fit returns a value for the random coincidences compatible with zero $(0\pm1)$, in agreement with the expectations. The best estimate for half-life of the decay results to be \tpo{}$= (143.3 \pm 2.8 )$~ms.
    The goodness-of-fit is satisfactory and the pull distribution (bottom) shows a good agreement between data and model.}
    \label{fig:Fit}
\end{figure}

The reconstructed time difference between \rnddz{}-\podus{} events is reported in Figure \ref{fig:Fit}. 
To extract an estimate of the \podus{} half-life (\tpo{}), we perform an unbinned extended maximum likelihood (UEML) fit with an exponential decay plus a flat background. 
This model takes into account both the signal and the possible background due to random coincidences. 
The latter is expected to be negligible, given the distinctive signature of  \rnddz{}-\podus{} decay chain and the very low rate of $\alpha$-events ($<1.7 \times 10^{-4}$ Hz/crystal).
We run the fit exploiting all the available data in the range (25~ms, 4000~ms).
The goodness-of-fit is confirmed evaluating the $\chi^{2}/\text{NDF} = 161.82/195 = 0.83$, which reflects the purity of the data sample acquired.
The flat background component is confirmed to be negligible, as the fit estimates $(0\pm1)$ background counts. 
This behaviour has been observed in all the delayed coincidence analysis performed in \cuz{}~\cite{Azzolini:2021yft}.
The estimated value for the process half-life is \tpo{}$=(143.3\pm2.8)$~ms.

Since the selection procedure is based on a well-defined experimental signature and all the uncertain data have been excluded from the analysis, the main systematic uncertainties we expect are related either to the fitting procedure or the extraction of the time differences between events.

We test possible dependence on the fit strategy omitting the constant background component and performing a binned extended maximum likelihood (BEML) fit, separately. In all these tests the fit results remain compatible with the presented one, with a maximum change in the central value below 0.1\%.

We evaluate the effects of threshold-related bias in the time distribution by repeating the fit increasing the lower threshold from 25~ms to 145~ms.
The fit result remains stable, both in expectation values and goodness-of-fit, evaluating the half-life as $(143\pm4)$~ms, a value compatible with the presented one. 
Since the only effect of excluding the lower time differences is an increased uncertainty due to the loss in statistics, this test proves that any possible bias related to the evaluation of the time-difference in \rnddz{}-\podus{} pulses is negligible with respect to the statistical uncertainty of the result. 
After these tests, we can state that our result is limited not by systematic uncertainties but by statistical ones.

\section{Conclusions}
\label{Sec:Conclusions}
In this work, we presented the measurement of the \podus{} half-life performed with \cuz{}. 
The characteristic signature of such decay allowed us to correctly identify the events of interest, providing a pure sample of reliable data.
Through a dedicated analysis of these events, we reconstructed the times of subsequent \rnddz{} and \podus{} decays, obtaining a spectrum of the time differences in a zero-background condition.
From this spectrum, we estimated the \podus{} half-life \tpo{}$=[143.3\pm2.8\text{(stat.})\pm0.01\text{(syst.})]$~ms.
This result is competitive with the measurements used to evaluate the recommended value, which are reported in Table~\ref{tab:216Po_HalfLife_Summary}.
Adding our independent evaluation and the recent result reported in Ref.~\cite{NADDERD2017119} to this ensemble, a new best estimate of ($144.3\pm0.6$)~ms can be calculated.
This result was possible by combining the excellent features of cryogenic calorimeters, the high radiopurity of \cuz{}, and the powerful techniques for event selection based on delayed coincidences \cite{Baccolo:2021odk,Azzolini:2021yft}.
Other measurements performed with \cuz{} already established the possibility of extending the scope of cryogenic calorimeters, investigating the nuclear dynamics behind the \bbd decay~\cite{Azzolini:2019yib} and the violation of conservation law~\cite{Azzolini:2019swx}. 
Similarly, this work opens a new path to precision measurements performed exploiting this experimental technique.
In particular, a Pb-based bolometer as the one used in Ref.~\cite{pattavina_radiopurity_2019} (appropriately contaminated with \radds{}) could be used to test the effect of a metallic environment at cryogenic temperatures on the \podus{} half-live~\cite{raiola_first_2007}.

\begin{table}[]
    \centering
    \resizebox{0.5\textwidth}{!}{
    \begin{tabular}{lr|c|c}
         {\bf Technique}& & {\bf Half-life [ms]} & {\bf Reference}\\
         
         \hline
         Ionisation Chamber& [1911] &  145 $\pm$ 15  & \cite{Moseley:1911}\\
         Geiger Counter &[1942] &  158 $\pm$ 8   & \cite{Ward:42}\\
         Si-surface Detector &[1963] &  145 $\pm$ 2   & \cite{DIAMOND1963143}\\
         Scintillation counting &[1978] & 155 $\pm$ 4   & \cite{Hey7:1978}\\
         Scintillating Crystal &[2003] &  144 $\pm$ 8   & \cite{Danevich:2003ef}\\
         Si-surface Detector &[2017] &  144 $\pm$ 0.6 & \cite{NADDERD2017119}\\
         Cryogenic Calorimeter &[2021]&  143 $\pm$ 3   & This work\\
         \hline
         Global Average &         &  144.3 $\pm$ 0.6 & \\
         Recommended Value -1 & &  145 $\pm$ 2 & \cite{Wu:2007cdl}\\
         Recommended Value -2 & &  148 $\pm$ 4 & \cite{216Po_LNHB_CEA}\\
    
    \end{tabular}
    }
    \caption{Summary of the \podus{} half-life measurements carried out since the early twentieth century. The first recommended value is quoted in the Nuclear Data Sheets, and calculated using Refs.~\cite{DIAMOND1963143,Danevich:2003ef}.
    The second recommended value is reported in the Table de Radionucleides of LNE – LNHB/CEA, and calculated using Refs.~\cite{Moseley:1911,Ward:42,DIAMOND1963143,Danevich:2003ef}.}
    \label{tab:216Po_HalfLife_Summary}
\end{table}

\section*{Acknowledgments}
This work was partially supported by the European Research Council (FP7/2007-2013) under contract LUCIFER no. 247115.
We thank M. Iannone for his help in all the stages of the detector assembly, A. Pelosi for constructing the assembly line, M. Guetti for the assistance in the cryogenic operations, R. Gaigher for the mechanics of the calibration system, M. Lindozzi for the cryostat monitoring system, M. Perego for his invaluable help in many tasks, the mechanical workshop of LNGS (E. Tatananni, A. Rotilio, A. Corsi, and B. Romualdi) for the continuous help in the overall set-up design. We acknowledge the Dark Side Collaboration for the use of the low-radon clean room. This work makes use of the DIANA data analysis and APOLLO data acquisition software which has been developed by the Cuoricino, CUORE, LUCIFER and \cuz collaborations. This work makes use of the Arby software for Geant4 based Monte Carlo simulations, that has been developed in the framework of the Milano -- Bicocca R\&D activities and that is maintained by O. Cremonesi and S. Pozzi. Mattia Beretta acknowledges support of the US Department of Energy Office of Science, Office of Nuclear Physics under award No. DE-00ER41138.

\bibliographystyle{model1-num-names}

\bibliography{main}

\end{document}